\documentstyle[12pt]{article}

\title{\begin{flushright}
{\normalsize NUC-MINN-99/15-T\\
November 1999 \\}
\end{flushright}
\vspace*{0.3in}
{\bf THE LAST EIGHT MINUTES OF A PRIMORDIAL BLACK HOLE}}
\author{{\bf Joseph 
Kapusta}\vspace*{0.1in}\\
 {\it School of Physics and Astronomy, University of Minnesota}\\
 {\it Minneapolis, MN 55455, USA}}

\date{}

\begin{document}

\maketitle

\begin{center}
Abstract
\end{center}

About eight minutes before a black hole expires it has a decreasing mass of 
$10^{10}$ g, an increasing temperature of 1 TeV, and an increasing luminosity of 
$7\times 10^{27}$ erg/s.  I show that such a black hole is surrounded by a 
quasi-stationary shell of matter undergoing radial hydrodynamic expansion.  The 
inner radius of this shell is bounded by ten times the Schwarzschild radius 
of $1.6\times 10^{-5}$ fm and has a temperature about one-tenth that of the 
black hole.  The outer radius, as defined by the photosphere, is about 
1000 fm, has a local temperature of 100 keV, and is moving with a Lorentz gamma 
factor of $10^7$.  Most of the emitted radiation is in photons with small 
amounts in gravitons and neutrinos. I calculate the instantaneous photon 
spectrum and then integrate it over the last eight minutes to obtain the energy 
distribution $dN_{\gamma}/dE = 4\pi m_{\rm P}^2/15E^3$ for $E >$ several TeV .     

\newpage

Hawking radiation from black holes \cite{Hawk} is of fundamental interest 
because it relies on the application of relativistic quantum field theory in the 
presence of the strong field limit of gravity, a so far unique situation.  It is 
also of great interest because of the temperatures involved.  A black hole with 
mass $M$ radiates thermally with a temperature
\begin{equation}
T_{\rm h} = \frac{m_{\rm P}^2}{8\pi M}
\end{equation}
where $m_{\rm P} = G^{-1/2} = 1.22\times 10^{19}$ GeV is the Planck mass.  (I 
use units in which $\hbar = c = k_{\rm B} = 1$.)  In order for the black hole to 
evaporate it must have a temperature greater than that of the present-day
black-body radiation of the universe of 2.7 K = 2.3$\times 10^{-4}$ eV.
This implies that $M$ must be less than $0.8\%$ of the mass of the Earth,
hence the black hole must 
have been formed primordially and not from a stellar collapse.  The black hole 
temperature eventually goes to infinity as its mass goes to zero, although once 
$T_{\rm h}$ becomes comparable to the Planck mass the semi-classical calculation 
breaks down and the regime of full quantum gravity is entered.  Only in two 
other situations are such enormous temperatures achievable: In the early 
universe ($T$ similarly asymptotically high) and in central collisions of heavy 
nuclei like gold or lead ($T = 500$ MeV is expected at the RHIC (Relativistic 
Heavy Ion Collider) just completed at Brookhaven National Laboratory and $T = 1$ 
GeV is expected at the LHC (Large Hadron Collider) at CERN to be completed in 
2005).  The spontaneously broken chiral symmetry of QCD gets restored in a phase 
transition/rapid crossover at a temperature around 160 MeV while the 
spontaneously broken gauge symmetry in the electroweak sector of the standard 
model gets restored in a phase transition/rapid crossover at a temperature 
around 100 GeV.  The fact that temperatures of the latter order of magnitude 
will never be achieved in a terrestrial experiment motivates me here to study 
the fate of primordial black holes during the final minutes of their lives when 
their temperatures have risen to 100 GeV and above.  The fact that primordial 
black holes have not yet been observed \cite{review} does not deter me in the 
least.
    
When $T_{\rm h} \ll m_{\rm e}$ (electron mass) only photons, gravitons, 
and neutrinos will be created with any significant probability.  These 
particles will not interact with each other but will be emitted into 
the surrounding space with the speed of light.  Even when $T_{\rm h} 
\approx m_{\rm e}$ the Thomson cross section is too small to allow the 
photons to scatter very frequently in the rarified electron-positron 
plasma around the black hole.  This may change when  
$T_{\rm h} \approx 80-100$ MeV when muons and charged pions are created in 
abundance.  At somewhat higher temperatures hadrons are copiously produced and 
local thermal equilibrium may be achieved, although exactly how is an unsettled 
issue.  Are hadrons emitted directly by the black hole?  If so, they will be 
quite abundant at temperatures of order 150 MeV because their mass spectrum 
rises exponentially (Hagedorn growth as seen in the Particle Data Tables 
\cite{PDG}).  Because they are so massive they move nonrelativistically and may 
form a very dense equilibrated gas around the black hole.  But hadrons are 
composites of quarks and gluons, so perhaps quarks and gluon jets are emitted 
instead?  These jets must decay into the observable hadrons on a typical length 
scale of 1 fm and a typical time scale of 1 fm/c.
Once the hadrons appear they may form an 
equilibrated gas around the black hole just as if they had been produced 
directly albeit with some time delay.  One can find arguments both for 
\cite{for} and against \cite{again} thermal equilibrium being maintained by the 
strongly interacting hadrons outside the Scharzschild radius.  Certainly this is 
a very difficult and open problem in quantum statistical mechanics.  Fortunately 
the answer is unambiguous for black hole temperatures greater than 
the electroweak scale of 100 GeV, as we shall see.

I assume that a primordial black hole is surrounded by a shell of 
hydrodynamically expanding matter in local thermal equilibrium when $T_{\rm h} > 
100$ GeV.  This assumption will be shown to be self-consistent.  A detailed 
description of how this situation comes to be is a difficult problem as 
discussed above and is not addressed in this paper.

The relativistic hydrodynamic equations describing an adiabatic, steady-state, 
spherically symmetric flow with no net baryon number or electric charge and 
neglecting gravity were derived by Flammang and Thorne \cite{flammang}.  The 
equations may be cast in the following form.
\begin{eqnarray}
4\pi r^2 v \gamma^2 T s(T) &=& L_{\rm in} = \,\,{\rm luminosity} \\
\gamma T &=& \,\,{\rm constant}
\end{eqnarray}
Here $r$ is the radial coordinate, $v$ is the local flow velocity with the 
associated Lorentz factor $\gamma$, $T$ is the local temperature, and $s(T)$ is 
the entropy density of the fluid.  The $L_{\rm in}$ is the luminosity flowing 
into the fluid.  It is generally less than the total luminosity of the black 
hole.  For example, gravitons will escape without scattering from the matter and 
so will not participate in the hydrodyamic flow.  These equations were derived 
in the context of accretion but, of course, they apply equally well to 
expansion.  They were used to describe gamma-ray bursts from neutron stars by 
Paczy\`nski \cite{bohdan}.  Given the luminosity these equations allow one to 
find the functions $v(r)$ and $T(r)$ algebraically without solving any 
differential equations.

A black hole has a Schwarzschild radius $R_{\rm h} = 2M/m_{\rm P}^2 =
1/4\pi T$.  Note that $\pi T_{\rm h} \cdot 2R_{\rm h} = 1/2$.  Roughly, the 
average thermal momentum of a massless particle times the diameter of the black 
hole is 1/2.  This is just a manifestation of the uncertainty principle 
applied to the creation of an excitation in a confined region of space.  The 
luminosity is
\begin{equation}
L = -\frac{dM}{dt} = \alpha(M) \frac{m_{\rm P}^4}{M^2} =
64 \pi^2 \alpha(T_{\rm h}) T_{\rm h}^2
\end{equation}
where $\alpha(M)$ is a function reflecting the species of particles available 
for creation in the gravitational field of the black hole.  It is generally 
sufficient to consider only those particles with mass less than $T_{\rm h}$; 
more massive particles are exponentially suppressed by the Boltzmann factor.
I use
\begin{equation}
\alpha = 2.011\times 10^{-8} \left[ 4200 N_0 + 2035 N_{1/2} + 835 N_1 + 95 N_2 
\right] \, .
\end{equation}
Here $N_s$ is the net number of polarization degrees of freedom for all 
particles with spin $s$.  The coefficients for spin 1/2, 1 and 2 were computed 
by Page \cite{Page} and for spin 0 by Sanchez \cite{Sanchez}.  In the standard 
model $N_0 = 4$ (Higgs), $N_{1/2} = 90$ (three generations of quarks and 
leptons), $N_1 = 24$ (SU(3)$\times$SU(2)$\times$U(1) gauge theory), and $N_2 = 
2$ (gravitons).  This assumes $T_{\rm h}$ is greater than the temperature for 
the electroweak gauge symmetry restoration \cite{perplex}.  Numerically 
$\alpha(T_{\rm h} > 100 \,{\rm GeV}) = 4.43\times 10^{-3}$.

The entropy of a black hole is given by the area formula.
\begin{equation}
S_{\rm h} = m_{\rm P}^2 \pi R_{\rm h}^2 = 4\pi \frac{M^2}{m_{\rm P}^2}
= \frac{m_{\rm P}^2}{16\pi T_{\rm h}^2}
\end{equation}
The entropy per unit time lost by the black hole $-d S_{\rm h}/dt$ is to be 
equated with that flowing through the matter, $4\pi r^2 v \gamma s$.  Taking the 
ratio of this entropy rate to the luminosity allows us to determine the constant 
in the hydrodynamic equation.
\begin{equation}
\gamma T = T_{\rm h}
\end{equation}
This result is independent of the function $\alpha(M)$.

The entropy density of weakly interacting massless particles is
\begin{equation}
s(T) = \frac{4\pi^2}{90} d(T) T^3
\end{equation}
where $d(T)$ is the number of bosonic degrees of freedom; fermions get counted 
with a weight of 7/8.  Using this in eq. (2) it is found that $v$ is a 
monotonically increasing function of $r$ and that its minimum possible value is 
$v_{\rm min} = 1/\sqrt{3}$.  This is a sensible value since that is the speed of 
sound of a weakly interacting gas of massless particles.  The corresponding 
radius and fluid temperature are $r_{\rm min} = 1.43 R_{\rm h}$ and
$T(r_{\rm min}) = 0.816 T_{\rm h}$.  The former assumes that gravitons and 
neutrinos do not contribute to the entropy of the fluid.  The actual radius at 
which the fluid may be considered as thermalized could very well be greater than
$r_{\rm min}$, but the fact that the latter is already greater than the 
Schwarzschild radius is eminently sensible.

The expanding fluid is in local thermal equilibrium when the mean free paths or 
thermalization lengths of the particles are less than the characteristic 
distance over which the temperature changes.  This characteristic distance is
$l = \gamma T |dr/dT|$ (the gamma takes into account the Lorentz transformation 
from the rest frame of the black hole to the rest frame of the fluid).  When $r$ 
is greater than a few times the Schwarzschild radius $v$ is already approaching 
the speed of light and $\gamma \gg 1$.  Then $r^2 T^2 = 360 
\alpha(T_{\rm h})/\pi d(T)$ and neglecting the slow variation of $d$ with $T$ 
(except near a first order phase transition)
\begin{eqnarray}
l &=& \frac{K T_{\rm h}}{T^2} \, , \nonumber \\
K &=& 6\sqrt{\frac{10}{\pi}\frac{\alpha(T_{\rm h})}{d(T)}} \, .
\end{eqnarray}
Above 100 GeV the numerical value of 
$K$ (leaving out gravitons and neutrinos) is 0.069.

In a relativistic plasma it is not sufficient to consider only 2-body reactions.  
Of prime importance in achieving and maintaining local thermal equilibrium are 
multi-body processes such as $2 \rightarrow 3$ and
$3 \rightarrow 2$, etc.  This has been well-known when calculating quark-gluon 
plasma formation and evolution in high energy heavy ion collisions \cite{klaus} 
and has been emphasized by Heckler in the context of black hole evaporation 
\cite{for}.  This is a formidable task in the standard model with its 16 species 
of particles.  Instead I require that the Debye screening length for each of 
the gauge groups in the standard model be less than $l$. The Debye screening 
length is the inverse of the Debye screening mass $m^{\rm D}_n$ where $n =1, 2, 
3$ for the gauge groups U(1), SU(2), SU(3).
Generically $m^{\rm D}_n \propto g_nT$ where $g_n$ is the gauge coupling 
constant and the coefficient of proportionality is essentially the square root 
of the number of charge carriers \cite{kapbook}.  For example, for color SU(3)
\begin{equation}
m^{\rm D}_3 = g_3 \sqrt{1+N_{\rm f}/6}\,T
\end{equation}
where $N_{\rm f}$ is the number of light quark flavors at the temperature $T$.  
The numerical values of the gauge couplings are: $g_1 = 0.344$, $g_2 = 0.637$, 
and $g_3 = 1.18$ (evaluated at the scale $m_Z$) \cite{PDG}.  So within a factor 
of about 2 I have $m^{\rm D} \approx T$.  This estimate makes sense when 
compared to two other measures.  The average energy of a massless particle is 
$3T$ and so $m^{\rm D}$ is about three times the thermal DeBroglie 
wavelength.  On the other hand Carrington and Kapusta \cite{tau} calculated the 
mean time between two-body collisions in the standard model for temperatures 
greater than the electroweak symmetry restoration temperature in the process of 
calculating the viscosity in the relaxation time approximation.  Averaged over 
all particle species in the standard model one may infer from that paper an 
average time of $3.7/T$.  Taking into account multi-body reactions would 
decrease that by about a factor of two to four.

The conclusion to be drawn from the above is that local thermal equilibrium 
should be achieved when $1/T < l$ which translates to the condition
$T < T_{\rm h}/15$.  Once thermal equilibrium is achieved it is not lost as $r$ 
increases because $l$ increases as $1/T^2$ whereas the Debye screening length 
increases more slowly as $1/T$.  The smallest radius at which local equilibrium 
is maintained is obtained from the condition on the temperature to be 
approximately $13 R_{\rm h}$.  This means that the thermalized matter 
surrounding the black hole is far enough away so that it doesn't affect the 
Hawking radiation whose calculation had assumed that the black hole radiated 
into a vacuum.

An amazing observation in experiments on central collisions of lead nuclei at a 
laboratory beam energy of 160 GeV per nucleon at the SPS at CERN is that the 
outgoing hadrons, such as nucleons, pions, and kaons, have spectra that appear 
to be very thermal \cite{qm}.  In those collisions, quantities like local 
temperature, pressure, energy density, and particle density vary significantly 
over a length scale of several fm.  For black hole temperatures greater than 100 
GeV one may easily calculate $l$ in that region of the expanding matter where it 
is in the hadronic phase, where the local temperature is on the order of 100 to 
160 MeV, and find that it is much greater than several fm.  Therefore thermal 
equilibrium is maintainted at least down to that range of temperatures.

Following Paczy\`nski's analysis of neutron star gamma-ray bursters 
\cite{bohdan} let us assume that the matter maintains thermal equilibrium all 
the way down to a local temperature less than the electron mass.  Gravitons and 
neutrinos have already escaped, leaving only photons and a rarified plasma of 
nonrelativistic electrons and positrons.  The photosphere is defined as that 
radius where the optical depth is unity \cite{photo}.  Interaction of the 
photons with the 
electrons and positrons proceeds through the Thomson cross section
$\sigma_{\rm T}$.  Defining $n_{\pm}$ as the total density of
non-relativistic electrons and positrons one has the Thomson mean free path for 
photons at temperature $T$.
\begin{equation}
\lambda_{\rm T} = \frac{1}{\sigma_{\rm T} n_{\pm}} =
\frac{3}{8\pi} \frac{1}{\alpha_{\rm EM}^2} \frac{m_{\rm e}^2}{n_{\pm}} =
34.1\, \mbox{\AA} \left(\frac{m_{\rm e}}{T}\right)^{3/2} \exp(m_{\rm e}/T)
\end{equation}
Using $d = 2$ at these temperatures, corresponding to the entropy being 
dominated by the photons, the condition $\lambda_{\rm T} = l$ determines the 
temperature $T_{\rm ps}$ of the photosphere.
\begin{equation}
\sqrt{\frac{T_{\rm ps}}{m_{\rm e}}} \exp(m_{\rm e}/T_{\rm ps}) =
\frac{T_{\rm h}}{9.2 \, {\rm GeV}}
\end{equation}
Thus the photosphere temperature decreases as the black hole temperature 
increases, roughly in a logarithmic manner.  Requiring $T_{\rm ps} < 100$ keV 
means that $T_{\rm h} > 680$ GeV.  For black hole temperatures greater than 
about 700 GeV essentially all of the luminosity is in the form of photons, with 
the gravitons emitted directly without rescattering, and the neutrinos having 
lost thermal contact at some very small radius and correspondingly high 
temperature.
 
The Lorentz gamma factor at the photosphere is determined by eq. (7) to be 
$\gamma_{\rm ps} = T_{\rm h}/T_{\rm ps}$.  When the black hole temperature is 1 
TeV the temperature of the photosphere is 100 keV and $\gamma_{\rm ps} = 10^7$.  
Let us now calculate the radiation from a surface moving at relativistic 
velocity.  The phase space density $f$ is invariant.  Hence
\begin{equation}
f({\bf p}, {\bf x}, t) = f({\bf p}', {\bf x}', t') =
\frac{1}{\exp(E'/T_{\rm ps}) - 1}
\end{equation}
where the unprimed variables refer to the black hole rest frame and the primed 
variables refer to the rest frame of the fluid.  We desire the flux in the black 
hole rest frame.  Using $E'=\gamma_{\rm ps} E (1-v_{\rm ps} \cos\theta)$ the 
spectral flux is
\begin{equation}
\frac{dF_{\gamma}}{dE} = \frac{E^3}{2\pi^2} \int_0^1 d(\cos\theta) \, \cos\theta 
\,f(E, \cos\theta) \, .
\end{equation}
The denominator of $f$ may be expanded in a Taylor series and integrated term by 
term.  The resulting series cannot be summed in terms of elementary functions.  
However, in the limit that $\gamma_{\rm ps} \gg 1$ and
$E \gg T_{\rm ps}/\gamma_{\rm ps}$ (a very small energy indeed!) one gets
\begin{equation}
\frac{dF_{\gamma}}{dE} = -\frac{E^2 T_{\rm h}}{2\pi^2 \gamma_{\rm ps}^2}
\ln\left[1-\exp\left(-E/2T_{\rm h}\right) \right] \, .
\end{equation}
Integration over all energy gives the total flux (energy/area/time).
\begin{equation}
F_{\gamma} = \frac{8\pi^2}{90} \gamma_{\rm ps}^2 T_{\rm ps}^4 =
\frac{8\pi^2}{90} \frac{T_{\rm h}^4}{\gamma_{\rm ps}^2}
\end{equation}
One may also integrate eq. (13) exactly to get the total flux; in the limit
$\gamma_{\rm ps} \gg 1$ one obtains the same result.  The corresponding 
expressions for each flavor of neutrino are
\begin{eqnarray}
\frac{dF_{\nu}}{dE} &=& \frac{E^2 T_{\rm h}}{2\pi^2 \gamma_{\nu}^2}
\ln\left[1+\exp\left(-E/2T_{\rm h}\right) \right] \, , \nonumber \\
F_{\nu} &=& \frac{7\pi^2}{90} \gamma_{\nu}^2 T_{\nu}^4 =
\frac{7\pi^2}{90} \frac{T_{\rm h}^4}{\gamma_{\nu}^2} \, .      
\end{eqnarray}
The temperature and Lorentz gamma-factor at which the neutrinos lose thermal 
contact will be addressed in another paper.

Equating the total luminosity of the black hole, excluding that carried by 
gravitons and neutrinos, with $4\pi R_{\rm ps}^2 F_{\gamma}$ determines the 
radius of the photosphere.
\begin{equation}
R_{\rm ps}^2 = \frac{180}{\pi} \frac{\alpha_{\rm eff}}{T_{\rm ps}^2}
\end{equation}
Here $\alpha_{\rm eff} = \frac{406}{435}\alpha$ with $\alpha$ evaluated at 
$T_{\rm h} > 100$ GeV: the standard model degrees of freedom are assumed to hold 
all the way up to the Planck temperature.  Starting with a black hole of 
temperature $T_{\rm h} > 100$ GeV the time it takes to evaporate/explode is
\begin{equation}
\Delta t = \frac{m_{\rm P}^2}{3 \alpha (8\pi T_{\rm h})^3} \, .
\end{equation}
This is also the characteristic time scale for the rate of change of the 
luminosity of a black hole with temperature $T_{\rm h}$.  It is many orders of 
magnitude greater than the time it takes a light signal to travel from the 
Schwarzschild radius out to the photosphere, thus justifying the
quasi-stationary assumption.  A black hole with temperature 1 TeV has a 
Schwarzschild radius of $1.57\times 10^{-5}$ fm, its photosphere is at
$R_{\rm ps} = 970$ fm, and has 464 seconds to live.

Knowing how the temperature, flow velocity and radius of the photosphere evolve 
with time allows us to determine the energy distribution of photons emitted 
during the final minutes of the black hole.  Integrating from $t = -\Delta t$ to 
$t = 0$ gives
\begin{equation}
\frac{dN_{\gamma}}{dE} = -\frac{45}{4\pi^5} \frac{\alpha_{\rm eff}}{\alpha}
\frac{m_{\rm P}^2}{E^3} \int_0^{E/2T_{\rm h}} dx \, x^4 \, \ln(1-e^{-x}) \, .
\end{equation}
Both this integrated distribution and the instantaneous one, 
$4\pi R_{\rm ps}^2 dF_{\gamma}/dE$, are insensitive to the precise location of 
the photosphere.  In the limit $E \gg 2T_{\rm h}$ eq. (20) becomes
\begin{equation}
\frac{dN_{\gamma}}{dE} = \frac{4 \pi}{15} \frac{m_{\rm P}^2}{E^3} \, .
\end{equation}
This energy distribution should be valid for any primordial black hole for 
photon energies much larger than 1 TeV = $10^{12}$ eV.  

The hot shell of matter surrounding a primordial black hole provides a 
theoretical testing ground rivaled only by the big bang itself.  In addition to 
the questions already raised, one may contemplate baryon number violation at 
high temperature and how physics beyond the standard model might be 
important in the last few minutes in the life of a primordial black hole.  
Experimental discovery of exploding black holes will be one of the great 
challenges in the new millenium.

\section*{Acknowledgements}

I am grateful to G. Amelino-Camelia for many discussions on Hawking radiation 
and to Paul Ellis, Larry McLerran and Yong Qian for comments on the manuscript.  
This work was supported by the US Department of Energy under grant
DE-FG02-87ER40328.


\begin{thebibliography} {99}

\bibitem{Hawk} S. W. Hawking, Nature (London) {\bf 248}, 30 (1974); Commun. 
Math. Phys. 
{\bf 43}, 199 (1975).

\bibitem{review} B. J. Carr and J. H. MacGibbon, Phys. Rep. {\bf 307}, 141 
(1998).

\bibitem{PDG} Particle Data Group: {\it Review of Particle Physics}, Eur. Phys. 
J. {\bf C3}, 1 (1998).

\bibitem{for} A. F. Heckler, Phys. Rev. D {\bf 55}, 480 (1997); Phys. Rev. Lett. 
{\bf 78}, 3430 (1997).

\bibitem{again} J. Oliensis and C. T. Hill, Phys. Lett. {\bf B143}, 92 (1984); 
J. H. MacGibbon and B. R. Webber, Phys. Rev. D {\bf 41}, 3052 (1990); J. H. 
MacGibbon and B. J. Carr, Astrophys. J. {\bf 371}, 447 (1991); F. Halzen, E. 
Zas, J. H. MacGibbon and T. C. Weekes, Nature (London) {\bf 353}, 807 (1991). 

\bibitem{flammang} K. S. Thorne and R. A. Flammang, Mon. Not. R. Astr. Soc. {\bf 
194}, 475 (1981); R. A. Flammang, {\it ibid}. {\bf 199}, 833 (1982).

\bibitem{bohdan}  B. Paczy\`nski, Astrophys. J. {\bf 308}, L43 (1986). 

\bibitem{Page} D. N. Page, Phys. Rev. D {\bf 13}, 198 (1976).

\bibitem{Sanchez} N. Sanchez, Phys. Rev. D {\bf 18}, 1030 (1978).  The 
coefficient was extracted from figure 5 of this paper to within about 5\% 
accuracy.

\bibitem{perplex} One may usefully contemplate whether the helicity-0 component 
of the W and Z bosons below the symmetry restoring temperature is to be treated 
as the third component of a free massive vector boson or as a free massive 
scalar particle on account of the Higgs mechanism.   If the vexing problem of 
quark/gluon versus hadron production has not attracted the curiosity of the 
reader this ought to.

\bibitem{klaus} K. Kinder-Geiger, Phys. Rep. {\bf 258}, 237 (1995).

\bibitem{kapbook} J. I. Kapusta, {\em Finite Temperature Field Theory}
(Cambridge University Press, Cambridge, England, 1989).

\bibitem{tau}  M. E. Carrington and J. I. Kapusta, Phys. Rev. D {\bf 47}, 5304 
(1993).

\bibitem{qm} See, for example, the proceedings of the Quark Matter series of 
international conferences, the most recent in print being: {\it Proceedings of 
Quark Matter `97}, Nucl. Phys. {\bf A638}, (1998).

\bibitem{photo} D. Mihalas, {\em Stellar Atmospheres}, 2nd ed. (Freeman, San 
Francisco, 1978). 

\end{thebibliography}
\end{document}